\documentclass[pra,twocolumn,showpacs,floatfix]{revtex4}
\usepackage{graphicx}
\usepackage{tabularx}
\usepackage{amsmath}
\usepackage{amsfonts}
\usepackage{amssymb}
\usepackage{natbib}
\newcommand{\be}{\begin{eqnarray}}
\newcommand{\ee}{\end{eqnarray}}
\begin{document}
\title{Atomic diffraction from nanostructured optical potentials}
\author{G. L\'{e}v\^{e}que, C. Meier, R. Mathevet, C. Robilliard, J. Weiner}
\affiliation{Laboratoire de Collisions, Agr\'{e}gats et R\'{e}activit\'{e}\\
UMR 5589 du CNRS et l'Universit\'e Paul Sabatier
\\Institut de Recherche sur les Syst\`{e}mes Atomiques et
Mol\'{e}culaires Complexes\\Universit\'{e} Paul Sabatier\\31062
Toulouse Cedex 4 France}
\author{C. Girard}
\affiliation{Centre d'Elaboration de Mat\'{e}riaux et d'Etudes
Structurales\\29, rue Jeanne Marvig, BP 4347\\31055 Toulouse Cedex
4 France}
\author{J. C. Weeber}
\affiliation{Laboratoire de Physique de l'Universit\'{e} de
Bourgogne\\9 avenue A. Savary, F-21078\\Dijon, France}
\date{\today}

\begin{abstract}
We develop a versatile theoretical approach to the study of
ultracold atom diffractive scattering from light-field gratings by
combining calculations of the optical near field generated by
evanescent waves close to the surface of periodic nanostructured
arrays together with advanced atom wavepacket propagation
techniques. Nanometric 1-D and 2-D arrays with subwavelength
periodicity deposited on a transparent surface and optically
coupled to an evanescent wave source exhibit intensity and
polarisation gradients on the length scale of the object and can
produce strong near-field periodic modulation in the optical
potential above the structure.  As a specific and experimentally
practical example we calculate the diffraction of cold Cs atoms
dropped onto a periodic optical potential crafted from a 2-D
nanostructure array. For an "out of plane" configuration we
calculate a wide diffraction angle($\simeq$ 2 degrees) and about
60\% of the initial atom flux in diffraction orders $\pm 1$, an
encouraging result for future experiments.
\end{abstract}
 \maketitle

\section{Introduction}

The prospect of manipulating neutral atoms and molecules by light
forces acting at nanometer scale lengths offers fascinating but
experimentally challenging possibilities in many areas of atomic
physics. Atom optics\,\cite{atomoptic}, atom
nanolithography\,\cite{lithography1,lithography2}, and atom
interferometry\,\cite{atomoptic} are prime examples. The use of
high-refractive-index dielectric or metallic nanometric objects to
produce subwavelength localized light field
distributions\,\cite{Girard-Dereux:1996,Girard-Joachim-Gauthier:2000,Greffet-Carminati:1997}
raises the possibility of "integrated atom optics" in which atoms
or molecules can be confined, guided or diffracted above
nanostructured surfaces fabricated to a designed
shape\,\cite{Dowling,Barnett,Burke}. Tailoring optical potentials
for atom control, and possibly Bose-Einstein condensates, is
analogous to the use of micromagnetic fields for a similar
purpose\,\cite{Weinstein,Muller,Cassetari,Hansel,Dekker,Hinds:a,Hinds:b,Reichel,
Zimmermann}. Atom diffraction from a transmission optical grating
provided one of the early examples of light-field atom
manipulation\,\cite{Pritchard}. A proposal for\,\cite{Cook:82} and
realization of\,\cite{Mlynek} a reflection light-field grating
quickly followed. These early developments stimulated many
subsequent experimental and theoretical studies, and a pertinent
review has recently appeared\,\cite{henkel}. Until now reflection
gratings for matter waves have been implemented by the formation
of one dimensional evanescent standing waves produced by
counterpropagating laser beams undergoing total internal
reflection through a glass
prism.\\

The approach we present here contrasts markedly with this earlier
work on standing waves of sinusoidal form.  We study diffractive
scattering of cold atoms from an evanescent field, spatially
modulated by an array of nanometric objects with high index of
refraction and subwavelength periodicity deposited on a glass
surface\,\,\cite{Kobayashi-Sangu-Ito-Ohtsu:2000,Balykin}. These
evanescent fields and their interaction with atoms will exhibit
several novel features. First, rather than a pure sinusoidal
evanescent standing wave, nanostructured periodic corrugation
generates fields containing higher-order harmonics and an
intricate polarization distribution. Second, spatial gradients of
field intensity and polarization at length scales well below the
diffraction limit interact with the external and internal atomic
degrees of freedom in ways that strongly depend on the geometry
and material of the nanostructures employed.  Third, the intensity
and polarization map above the nanostructure array also depends
strongly on the intensity and polarization of the exciting light
source.\\

\begin{figure}\centering \includegraphics[width=2.80in,angle=-90]{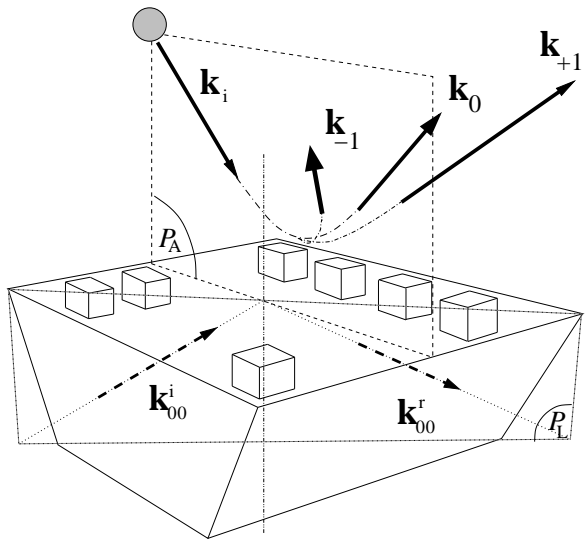}
\caption{Schematic diagram of cold-atom diffraction off an optical
potential.  The atom impinges on the periodic potential with
incident wave vector $\mathbf{k_i}$ and diffracts into orders
$\mathbf{k_{\pm 1},k_0}$. The periodic optical near-field above
the nanostructured array is generated by light with incident wave
vector $\mathbf{k^i_{00}}$, totally internally reflected into
$\mathbf{k^r_{00}}$. The nanostructures are cubes of
high-refractive-index material with subwavelength dimensions and
periodicity (see text for model details).  Note that the incident
plane of the light $P_L$ is along the diagonal of the cubes while
the incident plane of the atoms $P_A$ is shifted 45 degrees and
parallel to the cube edges.  Note also that the diffracted orders
$\mathbf{k_{\pm 1}}$ are "out-of-plane" with respect to the plane
of incidence of the light.} \label{fig0}
\end{figure}

Since the cold-atom de Broglie wavelength is not much smaller than
the characteristic scale length of the optical field, an accurate
description of atomic motion calls for a quantum treatment of
external as well as internal degrees of freedom. In order to
analyze the dynamics of cold atoms scattering off periodic optical
potentials we need to marry two well-developed numerical
techniques: calculation of the optical near-field and atom wave
packet propagation. The situation we consider is shown in
Fig.\,\ref{fig0}.
First, we calculate the three-dimensional electric field and
polarization distribution in the near field surrounding the
nanostructures. At frequency detunings far from resonance, where
absorption is negligible, this information is used to construct a
conservative potential for a full three-dimensional treatment of
the center-of-mass motion of the atom, including ground internal
states.  Using the light field information above the
nanostructures, we calculate the three-dimensional atom-field
interaction potential\,\cite{henkel,Henkel:97,Marani:00}. Since
the field polarization will vary significantly over the length
scale of the nanostructures \cite{leveque:01}, and since various
field polarizations may lead to population transfer among atomic
internal states\,\cite{henkel}, we include the internal Zeeman
states of a $^2S_{1/2}$ atomic ground level in the calculation of
the atom-optical field scattering.  Second, we apply a
time-dependent wave packet method to describe the scattering
problem of cold atoms diffracting from an optical grating with
subwavelength periodicity.  Inclusion of the ground-level internal
states leads to a three-dimensional, coupled channel problem that
we solve with wave packet propagation techniques already
successfully applied to atom or molecule surface scattering and
quantum molecular dynamics in several degrees of freedom
\cite{jackson,kosloff,uwe,Kroes:99,Lemoine:94a,Lemoine:92,Lemoine:97,mch:15,mch,ehara,Milot:98}.
After the wave packet representing the atom reflects and diffracts
from the optical potential, it is projected onto final scattering
states to yield the desired diffraction probabilities
\cite{mch:15}.\\

In this paper we emphasize general principles and seek to
establish a methodology without restriction to any specific
experimental set-up.  In fact the numerical solution of Maxwell's
equations in the near field and Schroedinger's equation for the
atomic motion can easily be adapted to explore atom manipulation
in sub-wavelength optical light fields of arbitrary geometry.
However to illustrate this methodology we present calculations
obtained with realistic parameters corresponding to an experiment
using a flux of cold $^{2}S_{1/2}$ atoms incident at about 40
degrees from the normal and scattering off a two-dimensional
optical grating with subwavelength period. The results of our
calculation yields out-of-plane diffraction into a few orders,
with markedly large diffraction angles. The results are
encouraging for planned
experiments.\\

The rest of the paper is organized as follows: Section II gives
details of the calculation of the scattering probabilities. The
sub-wavelength light field calculations are reported elsewhere
\cite{leveque:01}, so only the essential steps are summarized
here. In Section III we present a model configuration and in
Section IV we show the numerical results for illustration.

\section{Theory}

\subsection{The optical near field}

The evanescent wave field created by the totally internally
reflected laser beam is strongly modified by the periodic
nanostructures on the glass substrate surface. The calculation of
this intricate electric vector field above the nanostructures can
be carried out using several methods that solve Maxwell's
equations in these nontrivial geometries. These methods are well
established in the field of scanning near-field optical microscopy
(SNOM) \cite{Girard-Dereux:1996,Greffet-Carminati:1997}. Among the
most widely used are the finite-differences scheme
\cite{Kann:1995a}, the differential theory of gratings (DTG)
(\cite{Petit:1980,Maystre-Neviere:1978,Montiel-Neviere:1994}) and
the Green's function method\cite{Girard-Dereux:1996}. This latter
approach is well--adapted to study single, finite--size
nano--objects, and has been used extensively to study a wide range
of nanostructures
\cite{Girard-Dereux-Martin:1994,Martin-Girard-Smith-Schultz:1999}.
However, in the case of periodic surface structures, it becomes
inefficient, since it does not explicitly take the periodicity
into account. In such cases, the DTG method is more appropriate,
because it solves Maxwell's equations by means of a Fourier
expansion of each field component.

To be more specific, we define ${\bf r}=(x,y,z)$ with $z$ taken as the
direction perpendicular to the surface.  For convenience, we
define a vector ${\bf{l}}=(x,y)$ in the plane of the substrate
surface. We will denote the electric field by 1 in the glass half-space and
by 2 in the vacuum half-space, $n_1$ being the glass index and $n_2$ the vacuum index.
 The two-dimensional periodicity of the nanostructures
defines a unit cell of length $L_x$ and $L_y$ along the $x$ and
$y$ directions, respectively. The direction of the incoming laser
is given by $(k^x_L,k^y_L,k^z_L)$, with $k_L=n_12 \pi/\lambda_o$,
$\lambda_0$ being the laser wavelength in vacuum.

In the DTG method, an index-modulated zone, labelled 3 and characterized by $n_{3}({\bf r})$,
 the high index material, is introduced at the vacuum/glass substrate interface.
\begin{figure}[h!] \centering \includegraphics[width=1.9in,angle=-90.0]{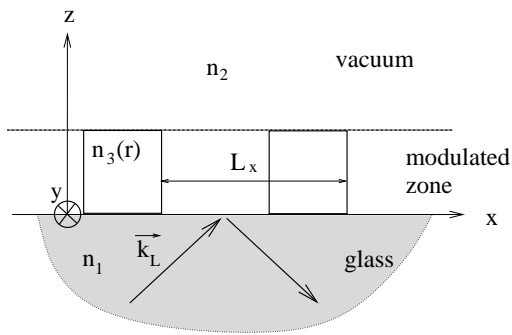}
\caption{Schematic of the three optical regions to which the
differential method of gratings is applied.} \label{figdtg}
\end{figure}
We have then to solve in all space Helmoltz's equation:
$$ \Delta {\bf E}^{(l)}({\bf r})+\mbox{n}^{2}_{l} k_0^2{\bf E}^{(l)}({\bf r})=0; \; l=1,2,3$$

Since the system is periodic along $x$ and $y$ we can expand the
dielectric constant (i.e. $n^{\,2}_{3}$) in the modulated zone:
$$ \mbox{n}^{\, 2}_{3}({\bf r}) = \sum_{n,m} \alpha_{mn} \exp(2 i\pi(\frac{n x}{L_x}+\frac{m y}{L_y})),$$
 where $\alpha_{mn}$ are the Fourier coefficient of $n^{\,2}_{3}$.
 It is clear that electric and magnetic fields {\emph{outside} the modulated
zone will also be periodic and thus conveniently written, \be
\label{DIFFRACTED} \mbox{\bf{E}}^{(l)}({\bf r}) & = & \sum_{m,n
=-\infty}^{\infty}\; \mbox{\bf{E}}^{(l)}_{mn}
e^{i\gamma_{mn}z}e^{i\mbox{\bf{k}}^\parallel_{mn}\cdot
\mbox{\bf{l}}} \\
\mbox{\bf{B}}^{(l)}({\bf r}) & = & \sum_{m,n = -\infty}^{\infty}\;
\mbox{\bf{B}}^{(l)}_{mn}
e^{i\gamma_{mn}z}e^{i\mbox{\bf{k}}^\parallel_{mn}\cdot
\mbox{\bf{l}}}; \; l=1,2 \ee In this expression,
$\mbox{\bf{k}}^\parallel_{mn}$ denote the $x$,$y$ components of
the wave vector  of the field diffracted in the $(m,n)$ order, and
is given by: $ \label{EYBY} \mbox{\bf{k}}^\parallel_{mn} = \left(
k^x_L + \frac{2\pi m}{L_x},k^y_L + \frac{2\pi n}{L_y}\right) $.
The $z$ component of the wave vector $\gamma_{mn}$ obeys the
dispersion equation: \be \label{KPAR}
\left(\mbox{\bf{k}}^\parallel_{mn}\right)^2 + \left(\gamma_{mn}
\right)^2 & = & \left(k^x_L\right)^2+ \left(k^y_L\right)^2+
\left(k^z_L\right)^2 \ee
 for which we have two solutions, corresponding
 to ``rising'' and ``descending'' waves. The rising zero order in the glass substrate
 corresponds to the incident laser field. Then, if we know the Fourier
components ${\bf E}_{mn}$ and ${\bf B}_{mn}$ of the electric and
magnetic field in a plane just above the nano--objects, the field
distribution anywhere above the structures are determined by
Eqs.\,\ref{DIFFRACTED},2.
 From Eq. \ref{KPAR}, it may be
seen that the coefficient $\gamma_{mn}$ is either real or
purely imaginary. The real values of $\gamma_{mn}$ correspond to
radiative harmonics while imaginary values introduce evanescent
components.

The six components of the electromagnetic field
$\mbox{\bf{E}}^{(l)}({\bf r})$, $\mbox{\bf{B}}^{(l)}({\bf r})$ are
then deduced from two independent parameters, usually named
{\emph{the principal components}, which in the present case are
chosen to be the $y$--components $E^{(l)y}({\bf r})$ and
$B^{(l)y}({\bf r})$. In order to calculate the Fourier components
of the electric field just above the nanostructures, we have to
solve a system of linear differential equations for $E^{(l)y}({\bf
r})$ and $B^{(l)y}({\bf r})$ in the region of space where the
index is modulated by the nanostructures. This system mixes all
the Fourier orders of the electric and the magnetic field through
the product $\mbox{n}^{\,2}_3E^{(3)y}({\bf r})$ in Helmholtz's
equation. Then, for numerical applications the Fourier expansion
converges at some sufficiently large value $N$, i.e. the order $n$
and $m$ will both vary
 from $-N$ to $+N$. We solve this system using
standard bounding conditions for this problem, i.e. we assume
there is no field arriving from infinity in the vacuum, and the
only Fourier components of the incident field in the glass
substrate are those of zero-order.  Finally, we can express the
Fourier components $E^{(2)y}_{mn}$ and $B^{(2)y}_{mn}$ in a
horizontal plane just above the nano-objects as a linear
combination of the $y$-components of the incident fields,
$E_i$,$B_i$, which define the polarization state of the incoming
laser : \be \label{coucou} E^{(2)y}_{mn}  & = &
{\mathcal{T}}^{EE}_{mn}\;E^{(1)y}_{i}
+ {\mathcal{T}}^{EB}_{mn}\;B^{(1)y}_{i} \\[0.3cm] B^{(2)y}_{mn}  & = &
{\mathcal{T}}^{BE}_{mn}\;E^{(1)y}_{i}
+{\mathcal{T}}^{BB}_{mn}\;B^{(1)y}_{i} . \ee  These transmission
coefficients,
${\mathcal{T}}^{EE}_{mn}$,${\mathcal{T}}^{EB}_{mn}$,${\mathcal{T}}^{BE}_{mn}$
and ${\mathcal{T}}^{BB}_{mn}$ depend only on the geometry of the
sample, the frequency and the angle of incidence of the
illuminating laser. The setup for the field calculation is shown
in  Fig.\,2.  A more detailed description of this calculation can
be found in Refs.\cite{Petit:1980,Montiel-Neviere:1994}.  With the
optical field mapping in hand we turn our attention to atom
scattering in the presence of these fields.
\\

\subsection{Calculation of atom-surface scattering}

In the limit of low saturation and a blue detuning $\delta$ which
is large compared to the Doppler shift and natural linewidth, the
atom-field interaction can be treated within the framework of
coherent atomic motion where spontaneous emission is neglected
\cite{henkel,Henkel:97,Marani:00}. We will consider an atom
transition dipole, typical of the first alkali $^2S_{1/2}
\rightarrow P$ transition, neglecting fine and hyperfine structure
in the excited state, but including the two-component angular
momentum degeneracy of the ground state. The excited level can be
eliminated adiabatically, which results in atomic motion that is
described by a two-component, three-dimensional wavepacket that
evolves on the ground-state manifold of the Zeeman sublevels
\cite{henkel,Henkel:97,Marani:00}: \be \label{schroed} i\hbar
\frac{\partial \Psi_{m_j}({\bf r})}{\partial t} & = & T_{{\bf r}}
\Psi_{m_j} + \sum_{m_j'=\pm \frac{1}{2}} V_{m_jm_j'}({\bf r}) \;\;
\Psi_{m_j'}({\bf r}) \ee In this expression, the operator of the
kinetic energy is simply given by \be \label{kinetic} T_{{\bf r}}
& = & -\frac{\hbar^{2}}{2 M} \nabla_{{\bf r}}^2 \ee with $M$ being
the mass of the atom. The potential within the low saturation
limit can be written as \cite{henkel,Henkel:97,Marani:00}
\begin{widetext}\be \label{potential} V_{m_jm_j'}({\bf r}) & = &
\frac{d^2}{\hbar \delta} \sum_{q,q',m_e} E^*_q({\bf r})
E_{q'}({\bf r}) (j_g,m_j;1,q|j_e,m_e)(j_e,m_e|j_g, m_j';1,q') \ee
\end{widetext} In this expression the terms in parenthesis are Clebsch-Gordan
coefficients with $j_g$ and $j_e$ being the total angular momentum
of the ground and excited state respectively. The reduced dipole
moment is denoted by $d$ and the field enters through its
spherical components $E_q({\bf r}), \;\; q=0, \pm1$.  This
expression forms the basis of numerous studies of atomic
diffraction from standing evanescent waves as has been used by
many authors \cite{Barnett,henkel,Henkel:97,Marani:00,Savage:95}.
In general the polarization state of the electric field above the
nanostructured objects is a very complicated function of space.
\begin{figure} \centering \includegraphics[width=3.25in]{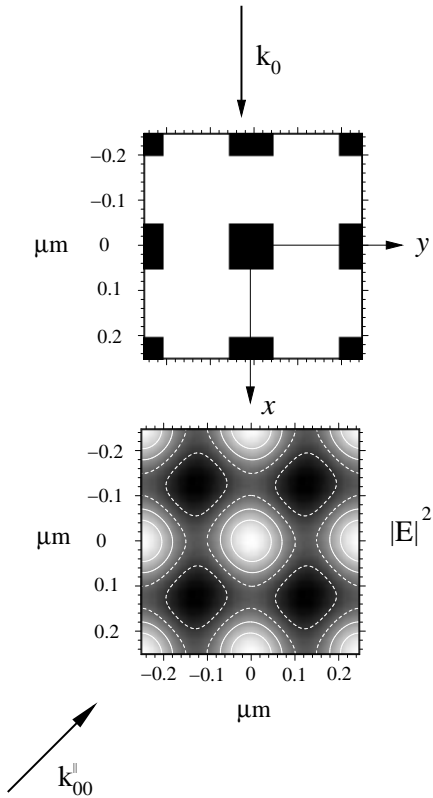}
\caption{Top panel: schematic view of the 100 nm TiO$_2$ cubes
deposited on a silica surface. Bottom panel: Light field intensity
distribution 125 nm above the nanostructures (the arrow
$\mbox{\bf{k}}^\parallel_{00}$ indicates the plane of the incident
laser of mode TM and ${\bf k}_i^{\parallel}$ the plane of incidence of the
atoms. The contour lines correspond to (in units of
$|E|^2_{inc}$): 0.397, 0.412 (dashed lines) and 0.427, 0.442 (full
lines)}\label{fig1}
\end{figure}
Hence, according to Eqs.\,\ref{schroed},\,\ref{potential},
transitions between the different Zeeman sublevels can occur
during the interaction of the atom with the light fields. As a
consequence, we performed the collision on the coupled ground
state levels.

One way of proceeding is to expand the wavefunction into plane
waves parallel to the surface which results in a set of coupled
diffraction channels for the direction perpendicular to the
surface\,\cite{henkel}. Diffraction then occurs as transitions
between these diffractive states, the probabilities of which can
be calculated by a semi-classical treatment within the
Landau-Zener theory,\cite{henkel} or by numerical wave packet
propagation\,\cite{Savage:95}. This treatment corresponds to the
well-known CCWP (close-coupling wave packet) method used in
standard atom or molecule surface scattering\,\cite{mch:10}.
However, with increasing diffraction orders one may need to take a
large number of diffractive states into account, together with all
the coupling matrix elements among them. In fact it is only
recently that modern and very efficient numerical wave packet
propagation techniques have permitted the calculation of
diffractive scattering by solving Eq.\,6 directly on grids in real
space\,\cite{mch,ehara,jackson}. Briefly, one simulates the
collisional process by a wave packet propagation from the initial
state to the final interaction--free zone, where it is projected
onto any desired observable, which in our case is the total
population in the different diffraction channels. To achieve a
good energy resolution, one needs to construct an initial wave
packet that is sufficiently large such that its energy width is
negligible. From a numerical point of view, this is very
disadvantageous, since it requires a large grid in the direction
perpendicular to the surface. Therefore, we used a method proposed
by Mowrey and Kouri\,\cite{mch:10}, who started with a spatially
narrow wave packet that comprised a wide range of energies, and
extracted energetically resolved results by projecting onto
asymptotic states of well-defined energy, ${\cal{E}}$.

The initial state of the atom is taken to to be a specific Zeeman sublevel $m^0$ of the
ground state manifold with center of mass motion described by the box-normalized wavefunction
\be \nonumber \label{psiinit} \lefteqn{\Psi^i_{m^0}({\bf r}) =
\left(2 \pi \xi^2 L_x^2 L_y^2 \right)^{-\frac{1}{4}} \times} \\
& & \exp \left({-(z-z_0)^2/4\xi^2 + i k^z_i z + i
\mbox{\bf{k}}_i\cdot \mbox{\bf{l}}} \right) \ee
with the initial transverse momentum $\mbox{\bf{k}}_i=(k^x_i,
k^y_i)$ and the initial momentum in $z$-direction described by a
Gaussian distribution centered around $k^z_i$. The propagation of
this three dimensional wavepacket is performed on the coupled
surfaces defined by Eqs. \ref{schroed},\ref{potential} until the
final scattered wave function $\Psi^f_{m_j}({\bf r})$ is entirely
in the asymptotic region.

The (unnormalized) modulus square of the transition amplitudes is
given by projection of the final wave packet onto diffractive
states\,\cite{mch:10}:
\be \nonumber \label{coeff} \lefteqn{|b^{mn}_{m_j}({\cal{E}})|^{2}  =}\\
& & \frac{1}{k^z_{mn}} \left| \int d\mbox{\bf{r}} \;\; e^{-i
\left(k^z_{mn} z + \left( \mbox{\bf{k}}_i+\mbox{\bf{G}}_{mn}
\right) \cdot \mbox{\bf{l}} \right)} \Psi^f_{m_j}({\bf r})
\right|^2 \ee
where $\mbox{\bf{G}}_{mn}=(\frac{2 \pi m}{L_x},\frac{2 \pi
n}{L_y})$ denotes the reciprocal lattice vector and $k^z_{mn}$ is
determined by energy conservation: \be \label{kz} k^z_{mn}=\sqrt{2
M {\cal{E}} - (\mbox{\bf{k}}_i+\mbox{\bf{G}}_{mn})^2}\ee The
probability to diffract into a specific order $(m,n)$ as a
function of total collision energy is then given
by\,\cite{ehara,mch:10}: \be \nonumber \label{Pmn}
\lefteqn{P_{mn}({\cal{E}})=} \\& & \left( \sum_{m_j=\pm
\frac{1}{2}} \left| b^{mn}_{m_j}({\cal{E}}) \right|^2 \right)
\bigg / \left( \sum_{m_j=\pm \frac{1}{2}} \sum_{mn} \left|
b^{mn}_{m_j}({\cal{E}}) \right|^2 \right) \ee This quantity will
be calculated for a selected set of parameters and discussed in
the next section. For the three-dimensional, two-component wave
packet propagation we used the FFT-Split-Operator scheme, which in
the field of quantum molecular dynamics or molecule-surface
collisions has proven to be a fast, stable and efficient method to
perform this task \cite{split}. Briefly, in this method the total
quantum mechanical short-time propagator is approximated by
\\
\begin{widetext} \be
\exp \left[ -i \frac{\Delta t}{\hbar} \left(
\begin{array}{cc}
T_{{\bf r}}+V_{\frac{1}{2},\frac{1}{2}}({\bf r}) &    V_{\frac{1}{2}, -\frac{1}{2}}({\bf r}) \\
V_{-\frac{1}{2},\frac{1}{2}}({\bf r})  & T_{{\bf r}}
+V_{-\frac{1}{2},-\frac{1}{2}}({\bf r})
\end{array}
\right) \right]
\approx \hspace*{5.5cm}\\[0.5cm]
\hspace*{-1cm} \exp \left[ -i \frac{\Delta t}{2 \hbar} \left(
\nonumber \begin{array}{cc}
T_{{\bf r}}  & 0 \\
0 & T_{{\bf r}}
\end{array}
\right) \right] \exp \left[-i \frac{\Delta t}{\hbar} \left(
\begin{array}{cc}
V_{ \frac{1}{2},\frac{1}{2}}({\bf r})  & V_{ \frac{1}{2},-\frac{1}{2}}({\bf r}) \\
V_{-\frac{1}{2},\frac{1}{2}}({\bf r})  &
V_{-\frac{1}{2},-\frac{1}{2}}({\bf r})
\end{array}
\right) \right] \exp \left[-i \frac{\Delta t}{2 \hbar} \left(
\begin{array}{cc}
T_{{\bf r}}  & 0 \\
0 & T_{{\bf r}}
\end{array}
\right) \right] \ee
\end{widetext}

This form of symmetric splitting is correct through second order;
it is unitary by construction and thus ensures numerical stability
\cite{split,FFTronnie}. The operator acts on the two-component
wavefunction to perform a short-time propagation from time $t$ to
$t+\Delta t$. The action of the kinetic operator is calculated in
Fourier space, where it is a simple multiplication with a phase
factor, and so every time step the two components of the
wavefunctions need to be Fourier transformed which can be done
very efficiently with three-dimensional FFT algorithms. With this
technique, the initial wavepacket Eq.(\ref{psiinit}) is propagated
until it is entirely in the asymptotic region. From which distance
the wavepacket can be considered to be effectively free has to be
checked carefully since it depends on many parameters. This is
especially important in the field of cold collisions where low
energies are considered.  Even though not applied in the current
approach, we note that the numerical effort can still be reduced
by an adiabatic correction of the initial state
\cite{mch,ehara,dieter}, analyzing the flux out of the scattering
region instead of projecting onto final diffractive states
\cite{dieter,Lemoine:94b} or using a filter-diagonalization scheme
\cite{filter}.

\section{Numerical Implementation}
\subsection{Model Configuration}

We model a typical experimental setup with a regular square
lattice of TiO$_2$ cubes (100 nm on a side, index of refraction
$n=2.1$) deposited onto a flat silica surface (index of refraction
$n=1.5$) at a center-to-center distance of 250 nm
(Figs.\,\ref{fig0},\ref{fig1},\ref{fig2}). These structures are
illuminated by an evanescent light field created by an incoming
laser with a vacuum wavelength of 850 nm, intensity of 80
W/cm$^{2}$, and subject to total internal reflection. The plane of
incidence of the laser beam (denoted
$\mbox{\bf{k}}^\parallel_{00}$ in Figs.\,\ref{fig0},\ref{fig1})is
chosen to be diagonal with respect to the rectangular
nanostructured pattern, and its polarization is taken to be within
the plane of incidence (TM polarization). The angle of incidence
with respect to the surface normal was chosen to be 60 degrees.
Under these conditions, the flat surface without the
nanostructures would give rise to an evanescent wave with a decay
length $\gamma_{00}$ of about 170 nm, and the atoms would not
approach closer than about 200 nm to the surface. Hence the
influence of the attractive van der Waals atom-surface potential
can be neglected. For different parameters however, it might
become important\,\cite{Landragin}, and can easily be included in
Eqs.\,\ref{schroed},\ref{potential}. The Fourier decomposition
Eq.\,\ref{DIFFRACTED} required $N=10$ terms for convergence.

\subsection{Diffractive scattering}

We have chosen a model atom with ground state $^{2}S_{1/2}$, the
mass of atomic Cs, and transition dipole moment corresponding to
the 6\hspace{0.1em}$^{2}S_{1/2}\rightarrow
6\hspace{0.1em}^{2}P_{3/2}$ atomic Cs transition.  The initial
internal state was taken to be $m_j=-1/2$ and the initial Gaussian
distribution of the perpendicular motion was centered around
$k_z^i=0.87$ nm$^{-1}$ with a width parameter $\xi=5.00$ nm. The
inital transverse momentum is taken to be $k_x^i=0.73$ nm$^{-1}$,
$k_y^i=0.0$. This corresponds approximately to cold Cs atoms
produced in a MOT after a free fall of about 1.5 cm before
colliding with the nanostructured surface inclined at an angle of
about 40 degrees between the vertical axis and the surface normal.
Under these conditions, the incoming and specular direction are in
the $y=0$ plane, as indicated in Fig.\,\ref{fig0}. We found
converged results for a cut-off distance of about 600 nm. The
asymptotic wavefunction is then analysed to yield the diffraction
probabilities $P_{mn}({\cal{E}})$ as outlined in the previous
section. In all calculations, grids of 16 points in the $x$ and 64
in the $y$ direction were used and 1024 points in the
$z$-direction.

\section{Results}

\subsection{Sub-wavelength optical near-field}

In Fig.\,\ref{fig1} we show the geometry of the TiO$_2$ cubes (a)
together with the intensity distribution (b) at a distance of 125
nm above the surface. The arrow indicating the plane of incidence
of the laser beam is denoted by $\mbox{\bf{k}}^\parallel_{00}$ and
plane of incidence of the atoms by $\mbox{\bf{k}}_i$. One can
clearly see that the light fields bear the periodicity of the
nanostructures. We found that illuminating the nanostructures with
the plane of laser incidence aligned along the cube diagonals
rather than along the sides yields periodic potentials with
steeper gradients (more pronounced localization) even at distances
far above the surface. At the illustrated distance of 125 nm above
the surface, we still find pronounced periodic field intensity
modulation. This strong localization of light intensity above
high-refractive-index structures is a well-known characteristic of
the TM illumination mode. It has been theoretically modelled and
experimentally verified by the SNOM technique for a number of
nanostructured objects \cite{Girard-Dereux:1996}. In
Fig.\,\ref{fig2} we show a cross section of the field intensity
distribution in the $y,z$ plane above the nanostructures up to 400
nm.
\begin{figure}\centering \includegraphics[width=3.25in, angle=-90]{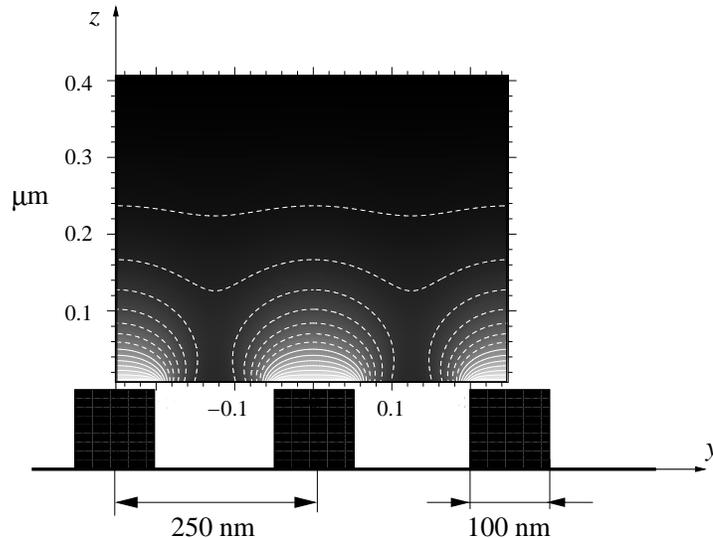}\caption{Light
field intensity distribution in the $y,z$ plane together with a
schematic view of the TiO$_2$ nanostructures deposited on the
silica surface. The contour lines correspond to (in units of
$|E|^2_{inc}$): 0.080, 0.135, 0.189, 0.241, 0.293, 0.346, 0.400
(dashed lines, from top to bottom).} \label{fig2}
\end{figure}
One clearly sees how the intensity contrast above the
nanostructures decays with increasing distance.  Depending on the
kinetic energy, the falling atoms will penetrate to different
heights above the surface, therefore experiencing a different
amplitude modulation (contrast) at the plane of the mean classical
turning point. It is important to note that even if the contrast
diminishes with distance, the periodicity remains the same; and
that even if the atoms are diffracted at large distances from the
surface, the diffraction angle will still be controlled by the
ratio of the atom de Broglie wavelength to the optical grating
period. In order to rigorously establish the optical potential
that governs the atomic motion, we need the full information of
the field above the surface, i.e. the three spherical components
of the electric field $E_+$, $E_-$ and $E_0$, as can be seen from
Eq. \ref{potential}. The quantization axis was chosen to be
perpendicular to the plane of laser incidence. From
Fig.\,\ref{fig3} one can see that the different components of the
light fields are a complicated function of space that will
interact with the multi-level internal structure of the colliding
atom.
\begin{figure}[t]\centering \includegraphics[width=3.25in]{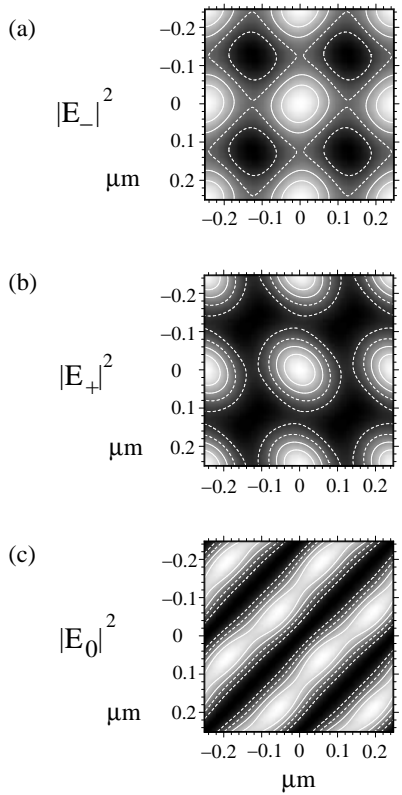}
\caption{Distribution of the spherical components of the electric
field 125 nm above the top of the nanostructures. The contour
lines correspond to (in units of $|E|^2_{inc}$): (a) 0.377, 0.387
(dashed lines) and 0.397, 0.407 (full lines): (b) 0.020, 0.025
(dashed) and 0.030, 0.035 (full lines): (c) 2.3$\cdot$10$^{-4}$,
4.6$\cdot$10$^{-4}$ (dashed) and 6.9$\cdot$10$^{-4}$,
9.2$\cdot$10$^{-4}$ (full lines).} \label{fig3}
\end{figure}
Note that the $E_0$ component is one order of magnitude smaller
than $E_+$ or $E_-$.

\subsection{Diffraction probabilities including ground state degeneracy}

In this section, we show the results for diffractive scattering,
using the light field parameters as detailed in the previous
section.  We stress that these probabilities are calculated using
the Cs 6\hspace{0.1em}$^{2}S_{1/2} \rightarrow
6\hspace{0.1em}^{2}P_{3/2}$ transition dipole and atomic mass, but
that no other specific atomic parameters were used. This particle
is still a model atom, however, since we have ignored hyperfine
structure in the ground and excited states. The center-of-mass
motion is treated entirely quantum mechanically, and the
polarization state of the electric field with its spatial
variation rigorously included in the calculations. Since we are
considering a 2-D potential surface, the diffraction takes place
in two spatial directions labelled $\mbox{\bf{G}}_{mn}$, which are
the reciprocal lattice vectors corresponding to the $x-$ and $y-$
directions, respectively.  The probability $P(m,n)$ to diffract
into a given order $(m,n)$ as a function of energy is given by Eq.
12. Figure \ref{fig4} plots these diffraction probabilities
$P(m,0)$ (top panel) and $P(0,n)$ (bottom panel) as a function of
total energy. The symmetrical results for negative values of $m,n$
are not shown for clarity. With the definitions employed $(m,0)$
corresponds to diffraction within the plane of incidence of the
atoms while $(0,n)$ corresponds to diffraction perpendicular to
this plane. These two cases will be called 'in- plane' and
'out-of-plane' diffraction, respectively.  Comparing the top and
bottom panels of Fig.\,\ref{fig4}, we see that probabilities for
in-plane and out-of-plane diffraction differ dramatically.

The top panel of Fig.\,\ref{fig4} shows the in-plane diffraction
probability as a function of energy for fixed initial transverse
momentum of $k^x_i=0.73$ nm$^{-1}$. For low energies, we find
negligible diffraction. Most of the final population is in the
(0,0) channel which represents specular reflection.
\begin{figure}\centering \includegraphics[width=3.25in]{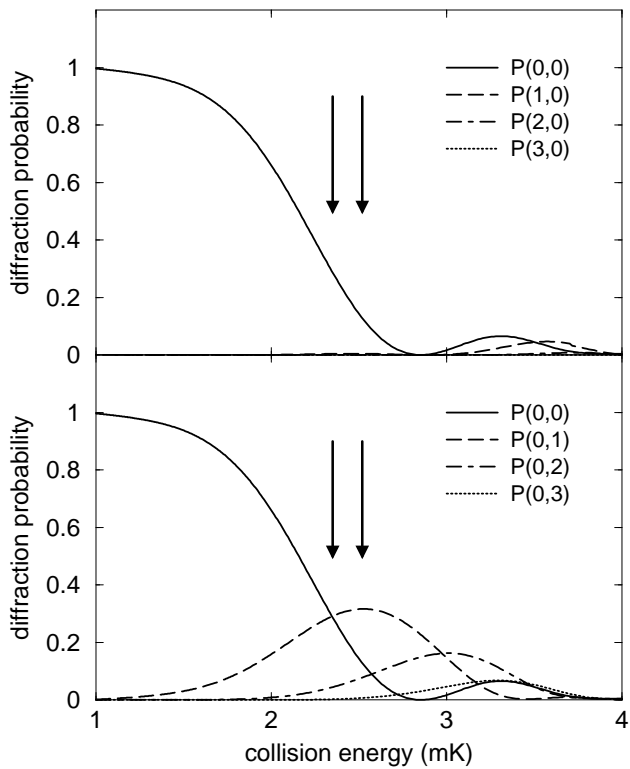}
\caption{Upper panel: In-plane diffraction probability $P(m,0)$
(the plane containing $\bf{k}_i$), Lower panel: Out-of-plane
diffraction probability $P(0,n)$ (perpendicular to the plane
containing $\bf{k}_i$), as a function of total collision
energy($E=k_{B}T$). The arrows indicate the cases discussed in the
text.} \label{fig4}
\end{figure}
As the energy $E$ increases, $P(0,0)$ decreases and falls off to
almost zero for energies greater than 2.5 mK ($E=k_BT$). The
higher orders of in-plane diffraction $P(m,0)$, $m=\pm 1,\pm 2,\pm
3$,.. are not significantly populated either. The bottom panel of
Fig.\,\ref{fig4} shows where the scattering flux has gone.  Almost
the entire population is found in the out-of-plane diffraction
orders ($0,n$), $n=\pm 1,\pm 2, \pm 3$. The fact that we calculate
a rapidly decreasing in-plane diffraction corresponds to the
well-known averaging of the transverse scattering amplitude of the
atomic motion parallel the diffracting structures \cite{henkel}.
With our parameters, we find an interaction time of about 3
$\mu$sec, which together with the initial transverse momentum of
$k^x_i=0.73$ nm$^{-1}$ implies that the atoms "sample" the
modulated potential over a transverse distance of about 800 nm,
greater than the periodicity of the potential by more than a
factor of three. For the out-of-plane scattering there is no such
averaging and the situation is analogous to in-plane diffraction
at normal incidence.  It is worthwhile noting that this
out-of-plane diffraction at grazing incidence (in a quite
different experimental arrangement than the one envisaged here)
has been observed experimentally \cite{brouri}.

Using the results shown in Fig.\,\ref{fig4}, one can now simply
read off the probability for specific collision energies. At an
energy of 2.35 mK, which corresponds to an angle of incidence of
about 40 degrees (indicated by the left arrows in the top and
bottom panels of Fig.\,\ref{fig4}), we have almost the entire
population equally distributed in the three diffraction orders
(0,0) and ($\pm 1$,0). Figure \ref{fig5} shows the complete set of
diffraction probabilities P($m,n$) for this case.
\begin{figure}\centering \includegraphics[width=3.25in]{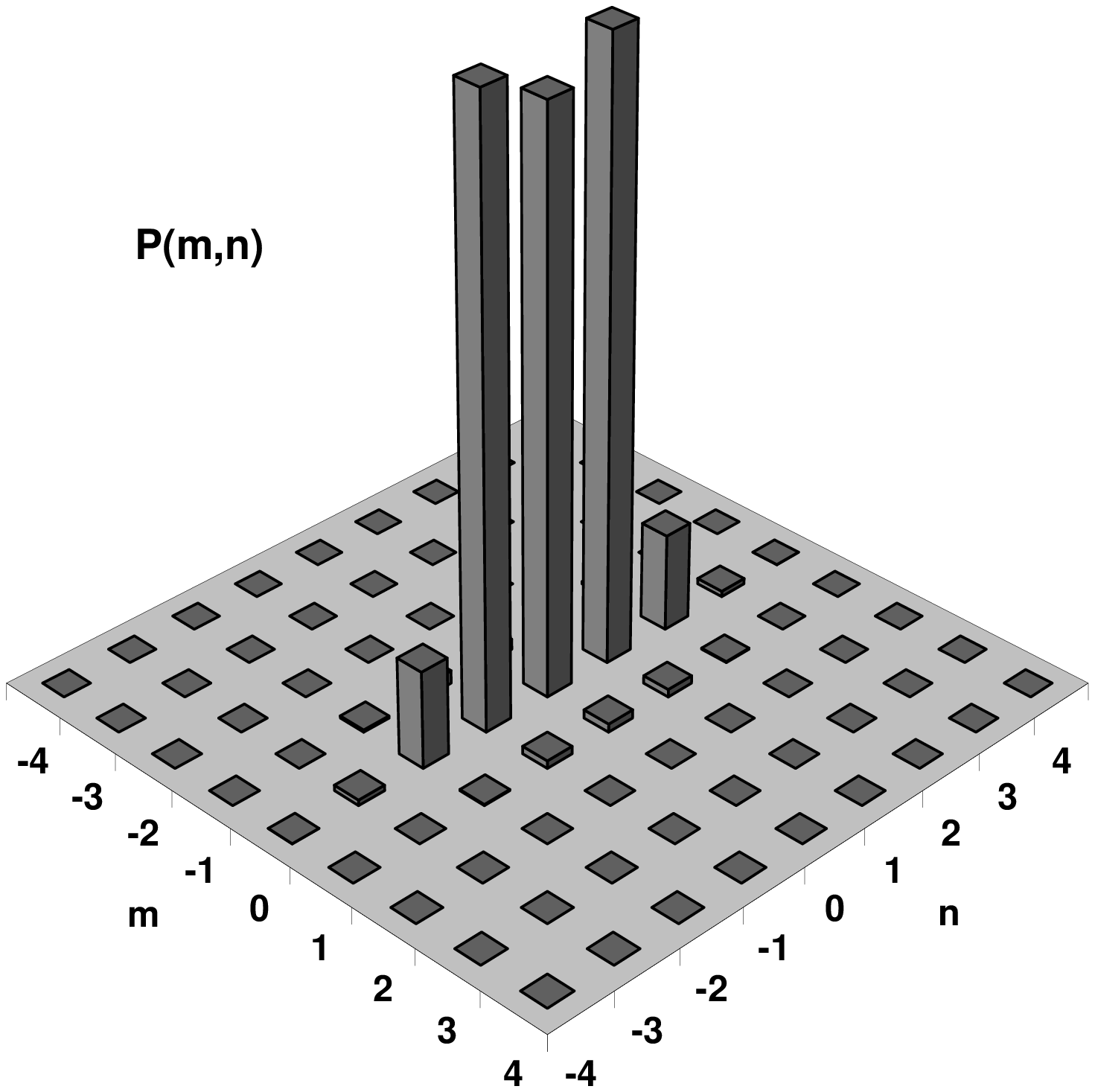}\caption{Diffraction
probabilities P($m,n$) for Cs atoms with a total energy of 2.35 mK
corresponding to a drop height of 15 mm (left arrows in Fig.\,3)
and an angle of incidence (between the vertical axis and the
surface normal) of 40 degrees.  The diffraction angle between the
orders ($0,\pm 1$) is $\simeq$ 2.6 degrees } \label{fig5}
\end{figure}
One sees clearly that the only significant diffraction is
out-of-plane, and the three dominating diffraction channels have
almost equal probability of 30 percent each. The diffraction
angle, given roughly by the ratio of deBroglie wavelength to
grating periodicity, is quite large--about 2.6 degrees. This
geometry could thus be a very attractive for the realization of an
atomic beam splitter or for atomic interferometry.

As a second example, we have shown in Fig.\,\ref{fig6} the
situation at slightly higher collision energy of 2.52 mK.
\begin{figure}\centering \includegraphics[width=3.25in]{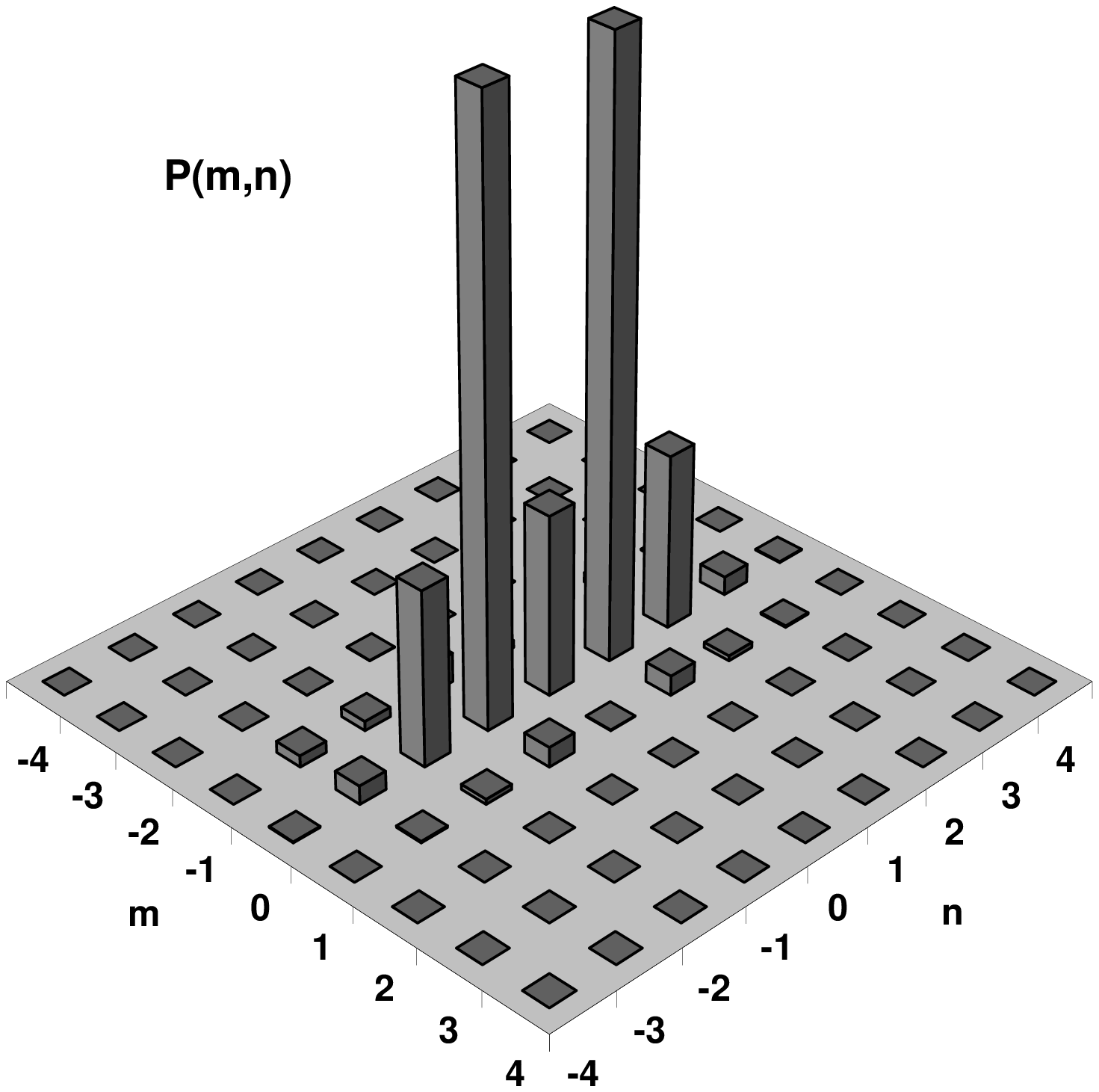}
\caption{Same as Fig.\,\ref{fig5}, but for a total energy of 2.56
mK, corresponding to a drop height of 16 mm,(right arrows in
Fig.\,\ref{fig4}) and an angle of incidence of 38 degrees.  The
diffraction angle between the orders ($0,\pm $)is $\simeq$ 2.0
degrees.} \label{fig6}
\end{figure}
With the chosen initial transverse momentum this corresponds to an
angle of incidence of about 38 degrees. The maximum of the final
state population is now concentrated in the orders (0,$\pm 1$)
while the specular direction is suppressed to about 10 percent. At
this slightly higher energy, we find also the diffraction peaks
(0,$\pm 2$) to be populated. Still, the ratio between P($0,0$) (or
P($0,\pm 2$)) and P($0,\pm 1$) is about 3, and the angle spanned
by the two main diffraction peaks P($0,1$) and P($0,-1$) is about
2 degrees.
\\

\subsection{Diffraction probabilities neglecting ground state degeneracy}

In the previous section, we have calculated  the diffraction
probability including the ground state degeneracy. As can be seen
from Eq.\,\ref{potential}, the intricate spatial distributions of
polarization in general do not allow treatment of the problem as a
one-surface scattering event, but require the solution of multiple
internal states coupled to the optical potential. However, as one
sees from Fig.\,3, the electric field is dominated by the $E_+$
and $E_-$ components, with the $E_0$ one order of magnitude
smaller (the axis for the polarization state is chosen to be
orthogonal to the plane of incidence of the laser beam). One sees
from Eq.\,\ref{potential} that the coupling between the ground
state sublevels in the present case of an $S_{1/2}\rightarrow
P_{3/2}$ transition is due to the $E_0$ component. Hence if the
component $E_0$ is sufficiently small, the system of Eqs.
\ref{schroed},\ref{potential} decouples and we have the situation
of an effective single-surface collision. Consistent with this
observation, we find that, starting from a well-defined initial
$m_j$ level, less than 0.1 percent of the population is
transferred to the other one. As a consequence, one can assume
that a treatment on one surface only, which greatly simplifies the
computational burden, can yield satisfactory results.  Thus in
this particular case a one-surface calculation gives very good
agreement with the full results, even for the weakly populated
diffraction orders. In general, however, with atoms having higher
ground-state multiplicity and with different optical materials and
excitation geometries, a full coupled-surface calculation will be
necessary.

\section{Conclusions}

In this paper we have investigated the diffraction of cold atoms
by highly structured subwavelength optical potentials generated
from evanescent fields. Our approach includes a three dimensional
quantum treatment of the atomic center of mass motion.  We take
into account the spatial distribution of rapidly varying
polarization states of the nanostructured optical fields and
include the effect of these polarization changes on the atom
ground internal state populations.

As an illustration, we have chosen a model system that corresponds
to cold cesium atoms (without nuclear spin) diffracting from a
nanostructured surface illuminated under conditions of total
internal reflection. The interaction of cold atoms with these
light fields is calculated in the limit of large detuning and
negligible absorption.  For experimentally realistic initial
conditions, we find diffraction angles on the order of 2 degrees
with about two thirds of the initial atomic flux concentrated in
the first two diffraction orders.  These structures may therefore
prove useful in wide-angle atomic interferometers.

\section{Acknowledgements}
Financial support from the Minist\`ere d'Education Nationale,
Recherche et Technologie, the Centre National de Recherche
Scientifique, the program Action Coordonn\'ee  Optique, and the
R\'egion Midi-Pyren\'ee is gratefully acknowledged.\\ Stimulating
discussions with D. Lemoine are also gratefully acknowledged.

\end{document}